\documentclass[ste,twoside]{stefano}
\usepackage{amsmath}
\usepackage{amssymb}
\usepackage{graphics}
\usepackage{rotating}
\usepackage{cite}
\usepackage{color}

\def\makeheadbox{{%
\hbox to0pt{\vbox{\baselineskip=10dd\hrule\hbox
to\hsize{\vrule\kern3pt\vbox{\kern3pt \hbox{  {\sc Eur. Phys. J. C
{\bf 46}, 551-558 (2006)} } \hbox{ {\sc
{\color{blue}{dma}}[{\color{black}{imecc}}]{\color{red}{UniCamp}}
} \hspace*{10.4cm} {\color{blue}{$\boldsymbol{\Sigma \delta
\Lambda}$}} }
\kern3pt}\hfil\kern3pt\vrule}\hrule}%
\hss}}}

\def\0{\mbox{\tiny $0$}}
\def\1{\mbox{\tiny $1$}}
\def\2{\mbox{\tiny $2$}}
\def\3{\mbox{\tiny $3$}}
\def\4{\mbox{\tiny $4$}}
\def\5{\mbox{\tiny $5$}}
\def\6{\mbox{\tiny $6$}}
\def\7{\mbox{\tiny $7$}}
\def\8{\mbox{\tiny $8$}}
\def\9{\mbox{\tiny $9$}}

\def\R{\mbox{\tiny $R$}}
\def\T{\mbox{\tiny $T$}}

\def\B{\mbox{\tiny $B$}}

\def\I{\mbox{\tiny $I$}}
\def\II{\mbox{\tiny $II$}}
\def\III{\mbox{\tiny $III$}}
\def\infm{\mbox{\tiny $-\infty$}}
\def\infp{\mbox{\tiny $+\infty$}}
\def\m{\mbox{\tiny min}}
\def\i{\mbox{\tiny inc}}

\def\mi{\mbox{\tiny $-$}}
\def\pl{\mbox{\tiny $+$}}

\begin{document}
%
%%%%%%%%%%%%%%%%%%%%%%%%%%%%%%%% PAPER %%%%%%%%%%%%%%%%%%%%%%%%%%%%%%%%%%%%%

\title{\Large  ABOVE BARRIER DIRAC MULTIPLE SCATTERING AND RESONANCES}
%\subtitle{}

\author{
%Alex E. Bernardini\inst{1}
Stefano De Leo\inst{1}
%\thanks{Partially supported by the FAPESP grant 99/09008--5.}
%\and
%Gisele C. Ducati\inst{1,2}
%\thanks{Supported by a CAPES PhD fellowship.}
%\and
%Celso C. Nishi\inst{2}
\and Pietro P. Rotelli\inst{2} }

\institute{
%Department of Cosmic Rays and Chronology, State
%University of Campinas\\
%PO Box 6165, SP 13083-970, Campinas, Brazil\\
%{\em alexeb@ifi.unicamp.br} \and
Department of Applied Mathematics, State University of Campinas\\
PO Box 6065, SP 13083-970, Campinas, Brazil\\
{\em deleo@ime.unicamp.br}
%{\em ducati@ime.unicamp.br}
%\and
%Department of Mathematics, University of Parana\\
%PO Box 19081, PR 81531-970, Curitiba, Brazil\\
%{\em ducati@mat.ufpr.br}
%\and Department of Cosmic Rays and Chronology, State University of
%Campinas\\
%PO Box 6165, SP 13083-970, Campinas, Brazil\\
%{\em ccnishi@ifi.unicamp.br}
\and
Department of Physics, INFN, University of Lecce\\
PO Box 193, 73100, Lecce, Italy\\
{\em rotelli@le.infn.it}
}

%%%%%%%%%%%%%%%%%%%%%%%%%%%%%%%%%%%%%%%%%%%%%%%%%%%%%%%%%%%%%%%%%%%%%%%%%%%
%%%%%%%%%%%% DATE ABSTRACT PACS % %%%%%%%%%%%%%%%%%%%%%%%%%%%%%%%%%%%%%%%%%

\date{Submitted: {\em November, 2005}. Revised: {\em January, 2006}.}
%- Revised version:  {\em April, 2004} }
% Warning: Where is the date?

\abstract{We extend an above barrier analysis made with the
Schr\"odinger equation to the Dirac equation. We demonstrate the
perfect agreement between the barrier results and back to back
steps. This implies the existence of multiple (indeed infinite)
reflected and transmitted wave packets. These packets may be well
separated in space or partially overlap. In the latter case
interference effects can occur. For the extreme case of total
overlap we encounter resonances. The conditions under which
resonance phenomena can be observed is discussed and illustrated
by numerical calculations.}

%%%%%%%%%%%%%%%%%%%%%%%%%%%%%%%%%%%%%%%%%%%%%%%%%%%%%%%%%%%%%%%%%%%%%%%
%%%%%%%%%%%%%%%%%%%%%%%%%%%%%%%%%%%%%%%%%%%%%%%%%%%%%%%%%%%%%%%%%%%%%%%

%%%%%%%%%%%%%%%%%%%%%%%%%%%%%%%%%%%%%%%%%%%%%%%%%%%%%%%%%%%%%%%%%%%%%%%
%%%%%%%%%%%%%%%%%%%%%%%%%%%%%%%%%%%%%%%%%%%%%%%%%%%%%%%%%%%%%%%%%%%%%%%

%\PACS{ {03.65.Xp}{}}

% Warning: No PACS code given

%02.10.Hh Rings and algebras
%02.10.Ud Linear algebra
%02.10.Yn Matrix theory

%02.30.Hq Ordinary differential equations
%02.30.Jr Partial differential equations
%02.30.Tb Operator theory

%03.65.-w Quantum mechanics
%03.65.Ca Formalism
%03.65.Ta Foundations of quantum mechanics; measurement theory
%03.65.Xp Tunnelling, traversal time,quantum Zeno dynamics
%03.65.Pm Relativistic wave equations

%12.15.F Quarks and lepton masses and mixing
%14.60.Pq Neutrino mass and mixing

%\offprints{~Stefano De Leo.}

\titlerunning{\sc dirac equation: multiple scattering and resonances}

\maketitle

%%%%%%%%%%%%%%%%%%%%%%%%%%%%%
%%%%%%%%%%%%%%%%%%%%%%%%%%%%%
\section*{I.  INTRODUCTION}
%%%%%%%%%%%%%%%%%%%%%%%%%%%%%
%%%%%%%%%%%%%%%%%%%%%%%%%%%%%

This paper considers the {\em above barrier} solutions of the
Dirac equation for an electrostatic one dimensional ($z$ axis)
potential,
 \[V(z) = \left\{
\begin{array}{rrl}
0\,\,, &   z < 0\,\, &\,\,\,\,\,(\mbox{\sc region I})\,\,,\\
V_{\0}\,\,, & \,\,\,\,0<z<l\,\, &\,\,\,\,\,(\mbox{\sc region II})\,\,,\\
0\,\,, &   \,\,\,z > l\,\, &\,\,\,\,\,(\mbox{\sc region III})\,\,.
\end{array} \right.
\]
It is very difficult and probably even confusing to treat, in a
single article, all interactions of plane waves or wave packets
with a barrier potential using the Dirac equation. This is because
the physical content depends upon the energy of the incoming
(particle) wave. In the figure below we depict the three potential
regions. Also shown is the energy divided into three zones.

\vspace*{.5cm}

\begin{picture}(180,90)
 \put(48,0){$0$}\put(141,0){$l$}
\put(63,70){\mbox{\small \sc Oscillatory}}
\put(63,62){\mbox{\small \sc (Particles)}}
\put(63,42){\mbox{\small \sc Evanescent}} \put(63,22){\mbox{\small
\sc Oscillatory}} \put(63,14){\mbox{\small \sc (Antiparticles)}}
 \put(13.7,71){\mbox{\small \sc I}}
 \put(13,64){\mbox{\small \sc n}}
 \put(13,57){\mbox{\small \sc c}}
 \put(13,50){\mbox{\small \sc o}}
 \put(13,43){\mbox{\small \sc m}}
 \put(14.5,36){\mbox{\small \sc i}}
 \put(13,29){\mbox{\small \sc n}}
 \put(13,22){\mbox{\small \sc g}}
\put(24.8,71){\mbox{\small \sc P}}
 \put(25,64){\mbox{\small \sc a}}
 \put(25,57){\mbox{\small \sc r}}
 \put(25,50){\mbox{\small \sc t}}
 \put(26,43){\mbox{\small \sc i}}
 \put(25,36){\mbox{\small \sc c}}
 \put(25,29){\mbox{\small \sc l}}
 \put(25,22){\mbox{\small \sc e}}
\put(25,15){\mbox{\small \sc s}}
 \put(167.3,71){\mbox{\small \sc O}}
 \put(168,64){\mbox{\small \sc u}}
 \put(168,57){\mbox{\small \sc t}}
 \put(168,50){\mbox{\small \sc g}}
 \put(168,43){\mbox{\small \sc o}}
 \put(169.5,36){\mbox{\small \sc i}}
 \put(168,29){\mbox{\small \sc n}}
 \put(168,22){\mbox{\small \sc g}}
\put(179.8,71){\mbox{\small \sc P}}
 \put(180,64){\mbox{\small \sc a}}
 \put(180,57){\mbox{\small \sc r}}
 \put(180,50){\mbox{\small \sc t}}
 \put(181,43){\mbox{\small \sc i}}
 \put(180,36){\mbox{\small \sc c}}
 \put(180,29){\mbox{\small \sc l}}
 \put(180,22){\mbox{\small \sc e}}
\put(180,15){\mbox{\small \sc s}} \put(5,-5){\mbox{\small \sc
region I}}\put(79,-5){\mbox{\small \sc region
II}}\put(160,-5){\mbox{\small \sc region III}}
 \thicklines
 \put(0,10){\line(1,0){50}}\put(140.5,10){\line(1,0){77}}
\put(50,10){\line(0,1){48}} \put(141,10){\line(0,1){48}}
\put(49.5,58){\line(1,0){92}}
 \put(0,57.){$....$}
\put(33,57.){$......$} \put(140.5,57.){$.........$}
\put(0,33){$....$}
\put(33,33){$................................................$}
\put(187,57.){$...........$}\put(187,33.){$...........$}
 \put(330,64){\mbox{\small \sc (Above
Barrier)}} \put(270,64){$E>V_{\0}+m$} \put(330,42,5){\mbox{\small
\sc (Tunnelling)}}
 \put(224,42.5){$V_{\0}-m <E<V_{\0}+m $}
 \put(330,18){\mbox{\small \sc
(Klein)}} \put(270,18){$E<V_{\0}-m$}
 \put(408.5,76){\mbox{\small \sc E}}
 \put(410,69){\mbox{\small \sc n}}
 \put(410,62){\mbox{\small \sc e}}
 \put(410,55){\mbox{\small \sc r}}
 \put(410,48){\mbox{\small \sc g}}
 \put(410,41){\mbox{\small \sc y}}
 \put(410,27){\mbox{\small \sc Z}}
 \put(410,20){\mbox{\small \sc o}}
 \put(410,13){\mbox{\small \sc n}}
\put(410,6){\mbox{\small \sc e}}
\end{picture}

 \vspace*{.5cm} \noindent
 The upper energy zone, $E>V_{\0} - m$, is that of interest to
 this work and involves diffusion phenomena.
 In the so called Klein zone\cite{K29}, $E<V_{\0} - m
$, oscillatory solutions exist in the barrier region. These are
antiparticles\cite{KLEIN1,KLEIN2,KLEIN3,KLEIN4,KLEIN5}. Indeed,
antiparticles see an opposite electrostatic potential to that seen
by the particles and hence they will see a well potential where
the particles see a barrier. The antiparticles thus live above the
well potential and are legitimately oscillatory in form. In the
tunnelling zone only evanescent waves exist in the barrier
region\cite{KSG01,PR03}. Of particular interest here is the
possibility of an Hartman-like effect\cite{H62,Rep,PLA}.

% A complete study of the barrier solutions of the Dirac equation
%can be divided quite naturally into three categories. The first,
%which is the subject matter of this paper, is limited to the
%oscillating solutions for above-barrier diffusion. This occurs for
%$E>V_{\0}+m$ where $V_{\0}$ is the barrier height (see the figure
%below, zone $A$). The second is actually made up of two energy
%zones: $V_{\0}+m>E>V_{\0}$ (zone $B$) where evanescent electron
%states exist in the classically forbidden region II ($0<z<l$) and
%$V_{\0}>E>V_{\0}-m$ (zone $C$) where antiparticle evanescent
%states exist in region II. These energy zones ($B$, $C$) exhibit
%tunnelling phenomena.  The third category is for $E<V_{\0}-m$
%(zone $D$). This is the Klein zone\cite{K29}. Technically, there
%is no tunnelling here since in region II only oscillatory
%solutions exist because $(E-V_{\0})^{\2}>m^{\2}$ even though
%$E-V_{\0}$ is negative. However, significant transmission into
%region III does occur. This is a consequence of the Klein paradox
%and its interpretation in terms of pair production...

No barrier analysis can be interpreted without an understanding of
the step potential results. In a latter Section, we shall argue,
for the step, that while pair production is innate in the Klein
zone there is no pair production for above barrier diffusion where
the non-relativistic limit reproduces the standard Schr\"odinger
solutions. There is of course no Schr\"odinger limit for the Klein
zone. The Klein zone of the step is characterized by the Klein
paradox in which a reflection probability greater than the
incoming probability occurs. The excess particle number (or
charge) implies pair production. For above step diffusion this
paradox does not exist and hence pair production is absent.

For the case of above barrier diffusion, we shall demonstrate in
detail the equivalence of a two-step calculation and the barrier
results. Such an equivalence has previously been demonstrated for
the Schr\"odinger equation\cite{MPLA}. This method of calculation
which employs the simple step results is not new. Multiple step
calculations have, for example, been used in the WKB applications
to various potentials\cite{AND} and for the barrier potential in
the Dirac equation in the Klein zone\cite{TK}. However, these
authors employ the method only as a convenient mathematical tool.
We, on the contrary, emphasize its physical significance in terms
of multiple reflected and transmitted peaks (when the barrier
width is much greater than the wave packets widths). The exit
times for each can be calculated using the stationary phase method
(SPM)\cite{KEL,COHEN}. They occur with time intervals equal to
twice the barrier width divided by the group velocity over the
barrier.

The combined summed expressions for  the reflection and
transmission coefficients contain the well known resonance
phenomena. We shall discuss under which conditions this resonance
effect occurs and when the effect "breaks up" and the multiple
peaks appear. Some numerical calculations will help us to
illustrate this transition from an effective single outgoing wave
packet (coherence) to essentially independent multiple wave
packets (de-coherence).

In the next Section we will give the necessary formulas and
conditions assumed for our calculations.
%While, not strictly
%necessary for this paper, we emphasize the electrostatic nature of
%our choice of potential. This means, that an antiparticle, such as
%that in the Klein zone, will "see" a potential-well and not a
%barrier.
In Section III, we consider the plane wave solutions for a step
potential and more specifically for three related but distinct
steps. One upward step at $z=0$ and two downward steps at $z=0$
and $z=l$. The extra phases that appear in the last case are
essential for the calculation of the times of the outgoing peaks.
We will then calculate the back to back step potential obtaining
the individual amplitudes for the (infinite) reflected and
transmitted terms. Each can be associated with a wave packet after
integrating with a suitable convolution function. In Section IV,
we calculate directly the plane wave solutions for the reflection
and transmission coefficients for a barrier and evidence their
equivalence to the sum of the results from the previous Section.
Section V discusses the question of resonance and points out that
this phenomena requires specific conditions to occur. We conclude
in Section VI with a resume of our results.

%%%%%%%%%%%%%%%%%%%%%%%%%%%%%%%%%%%%%%%%%%%%%%%%%%%%%%%%%
%%%%%%%%%%%%%%%%%%%%%%%%%%%%%%%%%%%%%%%%%%%%%%%%%%%%%%%%%
\section*{II.  DIRAC SOLUTIONS IN A CONSTANT POTENTIAL}
%%%%%%%%%%%%%%%%%%%%%%%%%%%%%%%%%%%%%%%%%%%%%%%%%%%%%%%%%
%%%%%%%%%%%%%%%%%%%%%%%%%%%%%%%%%%%%%%%%%%%%%%%%%%%%%%%%%

The free Dirac equation reads
\begin{equation}
\left( \,i\,\gamma^{\mu}\partial_{\mu} -
m\,\right)\Psi(\boldsymbol{r},t) = 0
\end{equation}
with the gamma matrices satisfying
$\left\{\,\gamma^{\mu}\,,\,\gamma^{\nu}\,\right\} = 2\,g^{\mu
\nu}$.  It has four well known independent plane wave solutions
 \[
u^{(\1,\2)}(\boldsymbol{p})\, \exp[\,i\,\boldsymbol{p} \cdot
\boldsymbol{r}-i\,E\,t \, ]\,\,,\,\,\,\,\,
(E=|E|)\hspace*{.5cm}\mbox{and}\hspace*{.5cm}
u^{(\3,\4)}(\boldsymbol{p})\, \exp[\, i\,\boldsymbol{p} \cdot
\boldsymbol{r}- i\,E\,t \, ]\,\,,\,\,\,\,\,(E=-|E|)\,\,,
\]
where $|E|=\sqrt{\boldsymbol{p}^{\2}+m^{\2}}$. Using the
Pauli-Dirac set of gamma matrices
\[ \gamma^{\0}=\left(\begin{array}{rr}\, 1 & 0\\
 0 & \,\,-1\end{array} \right)\hspace*{.5cm}\mbox{and}\hspace*{.5cm}
 \boldsymbol{\gamma}=
 \left(\begin{array}{cr} \,\,\,0 &\,\,\,\boldsymbol{\sigma}
  \\
 - \boldsymbol{\sigma} & 0\,\end{array} \right)\,\,,
\]
the spinorial solutions are (for polarized states in the
$z$-direction)
\begin{equation}
u^{(s)}(\boldsymbol{p})=N\,\left(
\begin{array}{c}
  \chi^{(s)}\\
\displaystyle{\frac{\boldsymbol{\sigma} \cdot
\boldsymbol{p}}{E+m}}\,\,\chi^{(s)}
\end{array} \right)\hspace*{.5cm}\mbox{and}\hspace*{.5cm}
u^{(s\pl  \2)}(\boldsymbol{p})=N\,\left(
\begin{array}{c}  -\, \displaystyle{\frac{\boldsymbol{\sigma}
\cdot \boldsymbol{p}}{|E|+m}}\,\,\chi^{(s)} \\
\,\,\,\,\,\,\chi^{(s)}
\end{array} \right)\,\,,
\end{equation}
where $s=1,2$ and
\[\chi^{(\1)}=\left( \begin{array}{c} 1 \\0\end{array} \right)
\hspace*{.5cm}\mbox{and}\hspace*{.5cm} \chi^{(\2)}=\left(
\begin{array}{c} 0 \\1\end{array} \right)\,\,,
\]
with $N = \sqrt{(|E|+m)}$ the covariant normalization choice such
that $ u^{(s)\,\dag} u^{(s)}=u^{(s\pl 2)\,\dag} u^{(s\pl
2)}=2\,|E|$. We recall that the free Dirac Hamiltonian is
\begin{equation}
H_{\0} = \boldsymbol{\alpha} \cdot \hat{\boldsymbol{p}} + \beta \,
m = \left( \begin{array}{cc} m & \boldsymbol{\sigma} \cdot
\hat{\boldsymbol{p}}\\  \boldsymbol{\sigma} \cdot
\hat{\boldsymbol{p}} & \,\,\,\,- m\,\,\,\,\end{array} \right)\,\,.
\end{equation}
The different signatures of the energies between $s$ and $s+2$
spinors follows readily from the alternative non-covariant form of
the Dirac equation
\begin{equation} i\,\partial_t
\Psi(\boldsymbol{r},t) = H_{\0}\Psi(\boldsymbol{r},t)
\end{equation}
 the left hand side yields
$E\,\Psi(\boldsymbol{r},t)$ in all four cases while the right hand
size yields $\pm \,|E|$ as the case may be. Actually, for
$\Psi^{(s)}(\boldsymbol{r},t)$ we appear to get a simple identity
because we have conventionally used $E$ rather than $|E|$ in its
spinor representation. However, when $\boldsymbol{p}\equiv
\boldsymbol{0}$ (the rest frame case) the Hamiltonian reduces to
$m\gamma^{\0}$ and the above equation yields $E=+\,m$ for $s=1,2$
and $E=-\,m$ for $s+2=3,4$. These solutions are oscillating
solutions, valid when $E\geq m$ or $E\leq -\,m$. There are also
evanescent solutions obtained from the above with the substitution
$\boldsymbol{p}\to i\, \boldsymbol{p}$ for which the spatial
dependence becomes $\exp[\mp\,\boldsymbol{p}]$ and occurs when
$-\,m<E<m$.

Since we shall need the solutions for step and barrier potentials,
we rewrite the above solutions in the presence of a constant
potential $V_{\0}$. Consider an {\em electrostatic potential}
$A_{\mu}=(A_{\0},\boldsymbol{0})$ included (via minimal coupling)
in the Dirac equation
\begin{equation}
\left( \,i\,\gamma^{\mu}\partial_{\mu} - e\, \gamma^{\0} A_{\0} -
m\,\right)\Psi(\boldsymbol{r},t) = 0
\end{equation}
or
\begin{equation}
i\,\partial_{t} \Psi(\boldsymbol{r},t) = (H_{\0} + V_{\0})\,
\Psi(\boldsymbol{r},t)
\end{equation}
where $V_{\0}=- \,e\,A_{\0}$ (charge $-\,e$). For a stationary
solution $\Psi(\boldsymbol{r},t) \propto \exp[-\,i\,E\,t]$, we
obtain
\begin{equation}
H_{\0} \Psi(\boldsymbol{r},t) = (E - V_{\0})\,
\Psi(\boldsymbol{r},t)\,\,.
\end{equation}
The spinorial solutions are now
\begin{equation}
u^{(s)}(\boldsymbol{q};V_{\0})=\tilde{N}\,
\left( \begin{array}{c}  \chi^{(s)}\\
\displaystyle{\frac{\boldsymbol{\sigma} \cdot
\boldsymbol{q}}{E-V_{\0}+m}}\,\,\chi^{(s)}
\end{array} \right)\,\, ,
\end{equation}
for $E - V_{\0} > m$, and
\begin{equation}
u^{(s\pl \2)}(\boldsymbol{q};V_{\0})=\tilde{N}\,\left(
\begin{array}{c} -\, \displaystyle{\frac{\boldsymbol{\sigma} \cdot
\boldsymbol{q}}{|E-V_{\0}|+m}}\,\,\chi^{(s)} \\ \,\,\,\,\,\,
\chi^{(s)}
\end{array} \right)\,\,,
\end{equation}
for $E - V_{\0} < -m$, where $(E-V_{\0})^{^{\2}}
-\boldsymbol{q}^{\2} = m^{\2}$ and $
 \tilde{N} =\sqrt{(|E-V_{\0}|+m)}$.
  This shows that the two set of solutions $u^{(s)}$ and $u^{(s\pl
\2)}$ are not determined by the sign of $E$ but by whether
$E>V_{\0}+m$ or $E<V_{\0}-m$, the latter being the defining
condition for the Klein zone. Hence, $E$ may be fixed but the
solutions depend upon whether in any given region the energy is
above or below the potential.

 In any constant potential region {\em only two}
solutions exist for a given $E$ (be they oscillatory or
evanescent). This fact is essential for a standard plane wave
analysis. Because the Dirac equation is a first order equation in
the spatial derivatives, for a step-wise continuous potential only
continuity of the wave function exists. There is no lose of
information w.r.t. the Schr\"odinger continuity (wave function and
its spatial derivative) since continuity of the "small" component
of the Dirac spinor yields in the non-relativistic limit the
continuity of the Schr\"odinger wave function's spatial
derivative. Thus, continuity of the Dirac wave function implies
four conditions at each interface, one for each spinor component.
However, since the {\em sign} of the momenta are a priori
arbitrary, we have two conditions which determine the spinor.
Hence, for consistency, there can be  {\em only two independent
spinors} in each region and these must correspond to different
helicity states or orthogonal combinations of them.

%%%%%%%%%%%%%%%%%%%%%%%%%%%%%%%%%%
%%%%%%%%%%%%%%%%%%%%%%%%%%%%%%%%%%
\section*{III. THE TWO STEP APPROACH}
%%%%%%%%%%%%%%%%%%%%%%%%%%%%%%%%%%
%%%%%%%%%%%%%%%%%%%%%%%%%%%%%%%%%%

Let us treat the barrier diffusions as an application of a two
step process. One upward step at $z=0$ and one downward step at
$z=l$. We assume an incoming above-potential, positive helicity,
plane wave state given by (the normalization of the spinor is
inessential here)
\[ \left(\,\begin{array}{c}1\\0\\ \displaystyle{\frac{p}{E+m}}\\0
\end{array}\,\right)\,\, \exp[\,i\,(p\,z-E\,t)\,]\,\,, \] with $E=\sqrt{p^{\,\2}
+m^{\2}}$. We shall need three step diffusions, the first at $z=0$
both for incoming waves  from the left and from the right. The
second one at $z=l$ only for  incoming waves from the left. For
reflection and transmission from an upward/downward potential we
shall use the suffixes  $+/-$.

\vspace*{.5cm}

\begin{picture}(480,90) \thinlines
\put(50,10){\vector(0,1){68}} \put(44,80){$V(z)$}
\put(102,31.5){$V_{\0}$} \put(2,10){\line(1,0){46}}
 \put(0,10){\vector(1,0){103}}
\put(106,8){$z$} \put(49,0){$0$} \put(10,19.5){\mbox{\small \sc
region I}} \put(56,19.5){\mbox{\small \sc region II}}
 \thicklines
\put(50,10){\line(0,1){25.5}} \put(50,35){\line(1,0){49}}
 \thinlines
\put(200,10){\vector(0,1){68}} \put(194,80){$V(z)$}
\put(203,31.5){$V_{\0}$} \put(152,10){\line(1,0){46}}
 \put(150,10){\vector(1,0){103}}
\put(256,8){$z$} \put(199,0){$l$} \put(157,19.5){\mbox{\small \sc
region II}} \put(206,19.5){\mbox{\small \sc region III}}
 \thicklines
\put(200,10){\line(0,1){25.5}} \put(200,35){\line(-1,0){49}}
 \thinlines
\put(350,10){\vector(0,1){68}} \put(344,80){$V(z)$}
\put(402,31.5){$V_{\0}$} \put(302,10){\line(1,0){46}}
 \put(300,10){\vector(1,0){103}}
\put(406,8){$z$} \put(349,0){$0$} \put(310,19.5){\mbox{\small \sc
region I}} \put(356,19.5){\mbox{\small \sc region II}}
 \thicklines
\put(350,10){\line(0,1){25.5}} \put(350,35){\line(1,0){49}}
\thicklines
\put(18,60){\vector(1,0){30}}
\put(52,50){\vector(1,0){30}}
 \put(48,50){\vector(-1,0){30}}
\put(12,57){$1$}
\put(-9,47){$R_{\pl}(0)$}\put(85,47){$T_{\pl}(0)$}
\put(168,60){\vector(1,0){30}} \put(202,50){\vector(1,0){30}}
 \put(198,50){\vector(-1,0){30}}
\put(162,57){$1$}
\put(143,47){$R_{\mi}(l)$}\put(235,47){$T_{\mi}(l)$}
\put(381.5,60){\vector(-1,0){30}} \put(352,50){\vector(1,0){30}}
 \put(348,50){\vector(-1,0){30}}
\put(384,57){$1$}
\put(291,47){$T_{\mi}(0)$}\put(385,47){$R_{\mi}(0)$}
\put(80,80){\mbox{\small \sc \fbox{step 1}}}
\put(230,80){\mbox{\small \sc \fbox{step 2}}}
\put(380,80){\mbox{\small \sc \fbox{step 3}}}
\end{picture}

\vspace*{.5cm}

\noindent For the first step, the solutions in region I and II are
respectively given by
\begin{equation}
\begin{array}{lcl}
 \Psi_{\I}(z,t) & = &
\left\{\, \left(\,\begin{array}{c}1\\0\\
\displaystyle{\frac{p}{E+m}}\\0
\end{array}\,\right)\, \exp[\,i\, p\,z\,] + R_{\pl}(0)\,
\left(\,\begin{array}{c}\,\,\,\,\,1\\\,\,\,\,\,0\\
\displaystyle{-\,\frac{p}{E+m}}\\\,\,\,\,\,0
\end{array}\,\right)\,
\,\, \exp[\,-\,i\, p\,z\,]\,\right\}\, \exp[\,-\,i\,
E\,t\,]\,\,,\\
  \Psi_{\II}(z,t) & = &
T_{\pl}(0)\,\left(\,\begin{array}{c}1\\0\\
\displaystyle{\frac{q}{E-V_{\0}+m}}\\0
\end{array}\,\right) \, \exp[\,i\, q\,z\,]\, \exp[\,-\,i\,
E\,t\,]\,\,,
\end{array}
\end{equation}
with $q=\sqrt{(E-V_{\0})^{^{\2}} -m^{\2}}$. All helicity (spin)
flip terms (the other independent spinor solutions) here and
elsewhere turn out to be absent, so we  exclude them a priori for
simplicity. Continuity at $z=0$ yields
\[
1 + R_{\pl}(0)  = T_{\pl}(0)\hspace*{.5cm}\mbox{and}\hspace*{.5cm}
 1 - R_{\pl}(0) =  T_{\pl}(0)\,
\frac{q\,(E+m)}{p\, (E-V_{\0}+m)} = T_{\pl}(0)\, \frac{b}{a}\,\,,
\]
where $a  = \sqrt{(E-V_{\0}+m)(E-m)}$ and $b =
\sqrt{(E-V_{\0}-m)(E+m)}$ ($a>b$). Thus, we obtain
\begin{equation}
\mbox{\sc Step 1 :}\,\,\,\left\{\,\begin{array}{rcl} R_{\pl}(0)&
=&(a-b)/\,(a+b)\,\,,\\  T_{\pl}(0) & = & 2\,a/\,(a+b)\,\,,
\end{array} \right.
\end{equation}
from which it follows that
\[
1-|R_{\pl}(0)|^{^{\2}} = \frac{b}{a}\,\, |T{\pl}(0)|^{^{\2}}\,\,.
\]
While $|R_{\pl}(0)|^{^{\2}}$ is obviously the reflection
probability, the "transmission" probability must therefore be
$(b/a)\,\, |T{\pl}(0)|^{^{\2}}$. A demonstration of this for wave
packets (albeit with the aid of some approximations) is given in
the Appendix. Since $|R_{\pl}(0)|<1$, we do not have a Klein
paradox here so no pair production is involved.

To understand physically the significance of single reflection
and/or transmission coefficients one needs to use the stationary
phase method for wave packets (normalized convolution integral of
plane waves).  This method,  introduced by Stokes and
Kelvin\cite{K29}, estimates the position of the maximum of the
wave packet by using the simple concept that, far from the
vanishing derivative of the phase, the argument of the convolution
integral oscillates many times and  produce destructive
interference. Consequently, the maximum of the wave packet occurs
where the derivative of the phase vanishes\cite{COHEN}. For
example, consider an amplitude
\[\Phi(z,t)=\int \mbox{d}k \,g(k)\, \exp[\,i\,\theta(k)\,]\, \exp[\,
i\,(\,k\,z-E\,t\,)\,]
\]
modulated by a real function $g(k)$, with a single steep maximum
at $k_{\0}$. The time-space relationship for the maximum (maxima
if $\Phi$ has spinor components) is given by
\[ t = \left(\,\frac{\mbox{d}k}{\mbox{d}E}\,\right)_{\0}z +
\left(\,\frac{\mbox{d}\theta}{\mbox{d}E}\,\right)_{\0}\,\,,
\]
the zero suffix means the parenthesis is calculated at $k=k_{\0}$.
The delay factor is $(\mbox{d}\theta/\mbox{d}E)_{\0}$. The group
velocity is $v_g= (\,\mbox{d}E/\mbox{d}k)_{\0}$. In our analysis
the {\em primary} modulation function, $g(p)$, is that of the
incoming wave packet
\begin{equation}
\Phi_{\I,\i}(z,t) =
\int_{p_{\m}}^{^{\infp}}\hspace*{-.2cm}\mbox{d}p\,\,\,
g(p)\,\left(\,\begin{array}{c}1\\0\\
\displaystyle{\frac{p}{E+m}}\\0
\end{array}\,\right)\, \exp[\,i\, (\,p\,z\,-E\,t\,)\,]\,\,.
\end{equation}
 The "effective" modulation function in any given region is then
given by $g\,A\,u_i$ where $A$ is the plane wave amplitude (e.g.
$R_{\pl}(0)$ or $T_{\pl}(0)$) and $u_i$ stands for the spinor
element considered. For each separate spinor component one must
calculate the group velocity and eventual delay times. However,
with our choice, all non zero spinor components are real, so that
there is no contribution to $\theta$ from them. Thus, with real
$R_{\pl}(0)$ and $T_{\pl}(0)$ we have no time delay. It is true
that the group velocities depends upon the spinor components
momentum dependence (which shifts the value of $k_{\0}$). However,
this should be negligible for a very sharply peaked modulation
function. It is demonstrably negligible in the two limits:
Non-relativistic (NR) where one can simply ignore the small
components; and in the ultra-relativistic (UR) when the ("small")
component $q/(E-V_{\0}+m)$ tends to one.

Returning to our calculations, we give without detail the results
for the other two steps. For step 2, we find
\begin{equation}
\begin{array}{ll}
\mbox{\sc Step 2 :}\,\,\, &\left\{\,\begin{array}{lcl}
R_{\mi}(l)&=&[\,(b-a)/\,(a+b)\,]\,\,\exp[\,2\,i\,q\,l\,]\,\,,\\
T_{\mi}(l)&=&[\,2\,b/\,(a+b)\,]\,\,\exp[\,i\,(\,q-p\,)\,l\,]\,\,,
\end{array}\right.
\end{array}
\end{equation}
and for step 3,
\begin{equation}
\begin{array}{ll}
 \mbox{\sc Step 3 :}\,\,\,&\left\{\,
\begin{array}{lcl}
R_{\mi}(0)&=&(b-a)/\,(a+b)\,\,,\\
T_{\mi}(0)&=&2\,b/\,(a+b)\,\,.
\end{array}\right.
\end{array}
\end{equation}
The first of the barrier transmitted amplitudes is thus obtained
by multiplying the "transmitted" amplitude of step 1,
$T_{\pl}(0)$, by the transmission amplitude of step 2,
$T_{\mi}(l)$,
\[ [\,4\,a\, b/\,(a+b)^{\2}\,]\,\,\exp[\,i\,(\,q-p\,)\,l\,]\,\,.    \]
 The exit time is calculated after including the
plane wave phase in region III,
\[ t = \left(\,\frac{\mbox{d}p}{\mbox{d}E}\,\right)_{\0}z +
\left(\,\frac{\mbox{d}\theta}{\mbox{d}E}\,\right)_{\0}\,\,,
\]
where $\theta=(q-p)\,l$. Thus, at  $z=l$ we find
\[t = \left[\, \left(\,\frac{\mbox{d}p}{\mbox{d}E}\,\right)_{\0} +
\left(\,\frac{\mbox{d}q}{\mbox{d}E}\,\right)_{\0} -
\left(\,\frac{\mbox{d}p}{\mbox{d}E}\,\right)_{\0}\right]\,l \,\,
 = \,
\left(\,\frac{\mbox{d}q}{\mbox{d}E}\,\right)_{\0} \,l \,\,.
\]
This is just the time for a wave packet in region II with  group
velocity $(\mbox{d}E/\mbox{d}q)_{\0}$ to travel a distance $l$
(barrier width). At step 2 there is also a reflected amplitude
given by $T_{\pl}(0)R_{\mi}(l)$. The corresponding wave packet
travels back towards $z=0$ and the second reflected wave exits
into the left region I, with amplitude (unmodulated)
$T_{\pl}(0)R_{\mi}(l)T_{\mi}(0)$ at the expected time
\[t=2\,(\mbox{d}q/\mbox{d}E)_{\0}\,l\,\,.\]
The whole procedure may then be repeated ad infinitum.

Below, we list the first few individual reflected and transmitted
waves together with the expressions for the general $n^{th}$ wave.
\begin{eqnarray}
R_{\1} & = & R_{\pl}(0)=\frac{a-b}{a+b}\,\,,\nonumber \\
R_{\2} & = &
T_{\pl}(0)\,R_{\mi}(l)\,T_{\mi}(0)=\frac{4\,a\,b\,(b-a)}{(a+b)^{\3}}\,
\exp[\,2\,i\,q\,l\,]
\,\,,\nonumber\\
R_{\3} & = &
T_{\pl}(0)\,R_{\mi}(l)\,R_{\mi}(0)\,R_{\mi}(l)\,T_{\mi}(0)=
\frac{4\,a\,b\,(b-a)}{(a+b)^{\3}}\, \exp[\,2\,i\,q\,l\,]
\left(\frac{b-a}{a+b}\,\exp[\,i\,q\,l\,]\right)^{\2}\hspace*{-.1cm}
,\nonumber \\
& \vdots & \nonumber \\ R_{n} & = &
T_{\pl}(0)\,R_{\mi}(l)\,[\,R_{\mi}(0)\,R_{\mi}(l)\,]^{n\mi
\2}\,T_{\mi}(0)=
\frac{4\,a\,b\,(b-a)}{(a+b)^{\3}}\,\,\exp[\,2\,i\,q\,l\,]\,
\left(\frac{b-a}{a+b}\,\exp[\,i\,q\,l\,]\right)^{\2 \,n
-\4}\hspace*{-.5cm}.
\end{eqnarray}
For the transmitted amplitude, we have
\begin{eqnarray}
T_{\1} & = &
T_{\pl}(0)\,T_{\mi}(l)=\frac{4\,a\,b}{(a+b)^{\2}}\,\exp[\,i\,(\,q-p\,)\,l\,]\,\,,
\nonumber \\
T_{\2} & = & T_{\pl}(0)\,R_{\mi}(l)\,R_{\mi}(0)\,T_{\mi}(l)=
\frac{4\,a\,b}{(a+b)^{\2}}\,\exp[\,i\,(\,q-p\,)\,l\,]\,
\left(\frac{b-a}{a+b}\,\exp[\,i\,q\,l\,]\right)^{\2}
\hspace*{-.1cm},\nonumber \\
T_{\3} & = &
T_{\pl}(0)\,[\,R_{\mi}(l)\,R_{\mi}(0)\,]^{\2}\,T_{\mi}(l)=
\frac{4\,a\,b}{(a+b)^{\2}}\,\exp[\,i\,(\,q-p\,)\,l\,]\,
\left(\frac{b-a}{a+b}\,\exp[\,i\,q\,l\,]\right)^{\4}\hspace*{-.1cm},\nonumber \\
& \vdots & \nonumber \\
T_{n} & = &
T_{\pl}(0)\,[\,R_{\mi}(l)\,R_{\mi}(0)\,]^{n\mi\1}\,T_{\mi}(l)=
\frac{4\,a\,b}{(a+b)^{\2}}\,\exp[\,i\,(\,q-p\,)\,l\,]\,
\left(\frac{b-a}{a+b}\,\exp[\,i\,q\,l\,]\right)^{\2\, n
-\2}\hspace*{-.5cm}.
\end{eqnarray}
At each step one can check (in accordance with our previous
discussion) that probability is conserved. {\em If the individual
wave packets are well separated} (see the following Section) the
probability of, say, the $n^{th}$ transmitted wave will be just
$|T_n|^{^{\2}}$ since it travels in a potential free region. A
straightforward calculation then shows that the total transmission
probability is,
\begin{equation}
\label{tn}
\sum_{n=1}^{\infty}\,|\,T_n\,|^{^{\2}}=\frac{16\,a^{\2}\,b^{\2}}{(a+b)^{\4}}\,
\left[\,1-\left(\displaystyle{\frac{b-a}{a+b}}\right)^{\4}\,\right]^{\mi
\1} = \frac{2\,a\,b}{a^{\2}+b^{\2}} \,\,.
\end{equation}
It is to be noted that this sum is independent of the barrier
width $l$. A similar calculation can be performed for the region I
reflected probabilities. This yields
\begin{equation}
\sum_{n=1}^{\infty}\,|\,R_n\,|^{^{\2}}=\left(\frac{a-b}{a+b}\right)^{\2}+
\frac{16\,a^{\2}\,b^{\2}\,(b-a)^{\2}}{(a+b)^{\6}}\,
\mbox{\Large{$/$}}\,
\left[\,1-\left(\displaystyle{\frac{b-a}{a+b}}\right)^{\4}\right]=
\frac{(a-b)^{\2}}{a^{\2}+b^{\2}} \,\,.
\end{equation}
Consequently, as expected
\begin{equation}
\sum_{n=1}^{\infty}\,\left\{\, |\,R_n\,|^{^{\2}} +
|\,T_n\,|^{^{\2}}\,\right\} = 1\,\,.
\end{equation}
Our multiple peak interpretation is thus consistent with overall
probability conservation. Finally, we observe that the time
interval $\Delta t$ between two successive outgoing peaks, in
either of the potential free regions is
\[\Delta
t=2\left(\,\frac{\mbox{d}q}{\mbox{d}E}\,\right)_{\0}l=2
\left(\,\frac{E-V_{\0}}{q}\,\right)_{\0}l\,\,.
\]

%%%%%%%%%%%%%%%%%%%%%%%%%%%%%%%%%%
%%%%%%%%%%%%%%%%%%%%%%%%%%%%%%%%%%
\section*{IV. THE BARRIER ANALYSIS}
%%%%%%%%%%%%%%%%%%%%%%%%%%%%%%%%%%
%%%%%%%%%%%%%%%%%%%%%%%%%%%%%%%%%%

Let us now perform the standard stationary plane wave analysis for
the barrier - again neglecting a priori (for simplicity)  spin
flip.

\begin{equation*}
\begin{array}{lclcl}
 \mbox{\small \sc Region I:} &~~~~ & \hspace*{.65cm} z < 0 \, ,& ~~~ &
 \hspace*{.85cm}
\left(\,\begin{array}{c}1\\0\\ \displaystyle{\frac{p}{E+m}}\\0
\end{array}\,\right)\,
 \exp[\,i\,p\;z\,] +  R\, \left(\,\begin{array}{c}\,\,\,\,\,1\\\,\,\,\,\,0\\
 - \displaystyle{\frac{p}{E+m}}\\0
\end{array}\,\right)\, \exp[\,-\,i\,p\;z\,]\,\,,
 \\
\mbox{\small \sc Region II:} & &0 < z < \,l \, ,&  &
\hspace*{.85cm} A\, \left(\,\begin{array}{c}1\\0\\
\displaystyle{\frac{q}{E-V_{\0}+m}}\\0
\end{array}\,\right)\,
 \exp[\,i\,q\;z\,] +  B\, \left(\,\begin{array}{c}\,\,\,\,\,1\\\,\,\,\,\,0\\
 - \displaystyle{\frac{q}{E-V_{\0}+m}}\\0
\end{array}\,\right)\, \exp[\,-\,i\,q\;z\,]\,\,,
\\
\mbox{\small \sc Region III:} &   & \,l < z \, ,&  &
\hspace*{.85cm} T\, \left(\,\begin{array}{c}1\\0\\
\displaystyle{\frac{p}{E+m}}\\0
\end{array}\,\right)\,
 \exp[\,i\,p\;z\,] \,\,.
\end{array}
\end{equation*}
Continuity at $z=0$ yields
\[ \left(\,\begin{array}{c}1\\R
\end{array}\,\right)=\frac{1}{2\,a}\,\left(\,\begin{array}{cr}a+b &
\,\,\,\,\,\,\,\,\,a-b\\a-b & a+b
\end{array}\,\right)\,\left(\,\begin{array}{c}A\\B
\end{array}\,\right)\,\,.
\]
From continuity at $z=l$,
\[\left(\,\begin{array}{cr}1 &
1\\1 & \,\,\,-1
\end{array}\,\right)\,\left(\,\begin{array}{l} A\,\exp[\,i\,q\,l\,]\\
B\,\exp[\,-\,i\,q\,l\,]
\end{array}\,\right)=\left(\,\begin{array}{c}1\\ a/\,b
\end{array}\,\right)\,T\,\exp[\,i\,p\;l\,]\,\,.
\]
Consequently,
\begin{eqnarray*}
\left(\,\begin{array}{l} A\\
B
\end{array}\,\right)&= & \frac{1}{2}\,
\left(\,\begin{array}{cc} \exp[\,-\,i\,q\,l\,] & 0\\
0 &\exp[\,i\,q\,l\,]
\end{array}\,\right)\,\left(\,\begin{array}{cr}1 & 1\\1 & \,\,\,-1
\end{array}\,\right)\,\left(\,\begin{array}{c}1\\ a/\,b
\end{array}\,\right)\,T\,\exp[\,i\,p\;l\,]\\ &=& \frac{\exp[\,i\,p\;l\,]}{2\,b}\,
\left[\,\begin{array}{l}(b+a)\,\exp[\,-\,i\,q\,l\,] \\ (b-a)\,
\exp[\,i\,q\,l\,]
\end{array}\,\right]\,T\,\,.
\end{eqnarray*}
Using this equation to eliminate $A$ and $B$ from the continuity
equation at $z=0$ gives
\[
\left(\,\begin{array}{c}1\\R
\end{array}\,\right)=\frac{\exp[\,i\,p\;l\,]}{4\,a\,b}\,
\left[\,\begin{array}{l}(a+b)^{\2}\,\exp[\,-\,i\,q\,l\,] -
(a-b)^{\2}\,\exp[\,i\,q\,l\,]
\\ (a^{\2}-b^{\2})\, \exp[\,-\,i\,q\,l\,] -(a^{\2}-b^{\2})\, \exp[\,i\,q\,l\,]
\end{array}\,\right]\,T\,\,.
\]
Whence,
\begin{eqnarray}
 R & = & i\, (b^{\2}-a^{\2})\,\sin[ql]\,\mbox{\Large$/$}
\left[\,2\,a\,b\,\cos[ql]-i\,(a^{\2}+b^{\2})\,\sin[ql]\,\right]\,\,.
\end{eqnarray}
and
\begin{eqnarray}
T&=&2\,a\,b\,\exp[\,-\,i\,p\;l\,]\,\mbox{\Large$/$}
\left[\,2\,a\,b\,\cos[ql]-i\,(a^{\2}+b^{\2})\,\sin[ql]\,\right]\,\,.
\end{eqnarray}
 These amplitudes satisfy,
\begin{equation}
|\,R\,|^{\2}+|\,T\,|^{\2}=1\,\,.
\end{equation}
If treated as
single wave packets (one for $R$ and one for $T$) the peaks emerge
at a {\em common} time determined by their common denominators
\[t_{\R}=t_{\T}=\left(\,\frac{\mbox{d}\theta}{\mbox{d}E}\,\right)_{\0}\,\,,
\hspace*{.5cm}  \mbox{with}\,\, \tan \theta =
\frac{a^{\2}+b^{2}}{2\,a\,b}\,\,\tan[\,q\,l\,]\,\,.
\]
Note that the phase factor $\exp[\,-\,i\,p\;l\,]$ in $T$ has been
cancelled by the plane wave factor $\exp[\,i\,p\;z\,]$ calculated
at $z=l$. The $i$ factor in $R$ is momentum independent so it does
not contribute to the time equation. Also for completeness, we
recall that $a$ and $b$ are real.

It is  always a little surprising that the two exit times
coincide. It also seems a little strange that the reflection time
delay ($t_{\R}$), compared to instantaneous reflection, depends
upon the barrier width $l$. Is the assumption of single reflection
and transmission peaks correct? This depends critically upon the
size/width  of the incoming wave packet. Before discussing further
what seems a result contrary  to the infinite multiple waves
described in the previous Section, we must make the following
important observation
\begin{equation}
T=\sum_{n=1}^{\infty}\,T_n
\hspace*{.5cm}\mbox{and}\hspace*{.5cm}
R=\sum_{n=1}^{\infty}\,R_n\,\,.
\end{equation}
Indeed while we have begun our analysis from the two-step
calculation we could have arrived at exactly the same result by
expanding the denominators of $R$ and $T$ in an infinite series.
The treatment of $R$ and $T$ as single wave packets represents the
limit situation in which all the infinite reflected wave packets
overlap and similarly for the transmitted wave packets. Plane
waves can in this sense be considered as infinitely extended wave
packets and they thus satisfy automatically this coherence
condition. The two approaches are perfectly equivalent. A single
peak may break up under suitable conditions into multiple peaks,
or equivalently, multiple peaks may coalesce under suitable
conditions into a single peak.

%%%%%%%%%%%%%%%%%%%%%%%%%%%%%%%%%%
%%%%%%%%%%%%%%%%%%%%%%%%%%%%%%%%%%
\section*{V. RESONANCE PHENOMENA}
%%%%%%%%%%%%%%%%%%%%%%%%%%%%%%%%%%
%%%%%%%%%%%%%%%%%%%%%%%%%%%%%%%%%%

One of the characteristic of the single wave packet situation is
manifest in the expressions for $R$ and $T$. The reflection
coefficient $R$ vanishes when $\sin[\,q\,l\,]=0$. It follows that
for values of $l$ such that
\[l=n\,\pi\,/\,q \hspace*{.7cm}\mbox{\small $(n=0,1,...)$}\,\,,\]
we obtain complete transmission. In the figure 1, we show,  for
different values of $V_{\0}$, $E_{\0}$ and $m$, the typical
resonance curves for $|T|$ (dotted lines).  In principle it
extends to infinite values of $l$. Any normalizing convolution
integral will modify this. First, an integration over $p$, unless
very "tight" about any $p_{\0}$ value will imply some averaging of
this curve. For a spread in momentum $\Delta p$ such that $\Delta
q \geq \pi\, q_{\0}$ the resonance effect will be completely
averaged out. Secondly, for an incoming wave packet with finite
spatial spread, say $\Delta z \sim 1/\Delta p$, we can ask when
the multiple peaks, described in the previous Section, are well
separated. This occurs when the distance between two peaks is much
greater than $\Delta z$. Thus, for complete decoherence
\[v_{g,\III}\,\Delta t \gg \Delta z \sim \frac{1}{\Delta p}\,\,,\]
where $v_{g,\III}$ is the group velocity in region III and $\Delta
t$ the time interval between successive peaks. Now $\Delta t =
2\,l/v_{g,\II}$, hence multiple peaks will be clearly separated
when
\[ 2\,l\, \frac{v_{g,\III}}{v_{g,\II}} \gg \Delta z
\hspace*{.5cm}\mbox{or}\hspace*{.5cm} \frac{2\,l\,}{\Delta z} \gg
\frac{v_{g,\II}}{v_{g,\III}}\,\,= \frac{q_{\0}}{p_{\0}}.
\]
For a plane wave $\Delta z=\infty$. Hence, for plane waves we have
maximum coherence always. For any finite ($\Delta z$), we see that
coherence is lost as $l$ increases. Decoherence implies (as proven
in the previous Section) that the total transmission probability
becomes independent of $l$ and is given by (\ref{tn}),
\[
\sum_{n=1}^{\infty}\,|\,T_n\,|^{^{\2}}=
\frac{2\,a\,b}{a^{\2}+b^{\2}} \,\,.
\]
This value happens to coincide with $|T|$ ({\em but not}
$|T|^{\2}$) at the mid resonance values where $\cos[ql]=0$. All of
this is exhibited in our numerical calculations, in the figure 1,
where the exact value of the transmission probability is plotted
against $l$. The tendency to a constant value as decoherence sets
in is apparent.

The condition for decoherence is obviously achieved as $l\to
\infty$. It is also obtained if $\Delta z\to 0$. However, $\Delta
z\to 0$ implies $\Delta q\to \infty$ and we must be careful not to
drop below the A (above barrier) zone. There is also a third limit
in which it occurs, when $\frac{v_{g,\II}}{v_{g,\III}}\to 0$.
 Let us now consider this last limit in more detail. It can be
achieved in two different ways. The first is by sending
$q_{\0}\to\ 0$ with $V_{\0}$ fixed, whence $p_{\0}\to
\sqrt{V_{\0}(V_{\0}+m)}$. The second is by keeping $q_{\0}$ fixed
and sending $p_{\0}(E_{\0})$ and $V_{\0}$ simultaneously to
infinity so that $E_{\0}-V_{\0}$ remains constant. The first
choice is again difficult to realize because of the Heisenberg
uncertainty principle. To see this concretely,  consider a {\em
symmetric} convolution function about $q_{\0}(p_{\0})$. As
$q_{\0}\to 0$ so to must $\Delta q \to 0$ since we must stay above
the tunnelling zone. This means that automatically we must have
$\Delta z \to \infty$. So contrary to our expectations we end up
in the coherent state. The second choice however does indeed lead
to complete decoherence since it can be achieved while keeping the
widths of the wave packets fixed.

%%%%%%%%%%%%%%%%%%%%%%%%%%%%%%%%%%
%%%%%%%%%%%%%%%%%%%%%%%%%%%%%%%%%%
\section*{VI. CONCLUSIONS}
%%%%%%%%%%%%%%%%%%%%%%%%%%%%%%%%%%
%%%%%%%%%%%%%%%%%%%%%%%%%%%%%%%%%%

In this paper we have considered diffusion of an incoming wave (or
wave packet) with $E>V_{\0}+m$ by a one-dimensional potential
barrier of height $V_{\0}$ and width $l$. In front and beyond the
barrier, the potential is assumed to be zero (free space). In this
study, we have employed the SPM. There is an inherent ambiguity in
this method. Given a sum of terms, it may be applied individually
to each or to the sum. In the former case a series of wave packet
peaks are determined while in the latter case only one peak is
predicted. It is easy to use the SPM in both the limit of complete
coherence and complete decoherence. It is difficult to see how to
use it for intermediate situations. In these cases we can fall
back upon pure numerical calculations or possible use cluster
decompositions in which the wave packets are summed in finite
numbers before applying the SPM. However, this possibility and its
viability has still to be explored.

The overall reflection and transmission amplitudes ($R$ and $T$)
are characterized by resonance oscillations and by the feature
that the reflection delay time is equal to the transmission time;
at least when (or to extent) that we can consider each a single
wave packet. What we have shown in this paper is that, even with
the Dirac equation, the barrier results can be obtained by
treating the barrier as a two step process. This procedure
involves multiple reflections at each "step" and when $l\to
\infty$ predicts the existence of multiple (infinite) outgoing
peaks. However, by simply summing the individual amplitudes, one
obtains exactly the standard barrier results. This has lead us to
postulate and then confirm numerically that with increasing $l$
the resonance curves will lose coherence and tend to predicted
constant values.

From an alternative, but equivalent viewpoint, it has also been
noted that, whereas one cannot perform the limit $l\to \infty$ in
the $R$ and $T$ amplitudes, one can expand the denominators, in a
natural way, into an infinite series which reproduces exactly the
two step results. For any finite but sufficiently large $l$ we
predict the appearance of multiple peaks of which the first
reflected term is simply the single step result characterized by
"instantaneous" reflection. As $l$ increases the exit times of the
other peaks grow. This suggests  that for {\em finite times} (or
simply ignoring secondary peaks) the single step is equivalent to
a barrier with a {\em sufficiently} large width. So not only can
we claim to have shown that a barrier is equivalent to two steps
but, at least for the first reflected wave, that a wide barrier is
in its turn an approximation of a single step.

We shall call upon these results in a subsequent work or works in
which we consider the tunnelling energy zones and the Klein zone.
We consider these energy zones separately because they are
characterized by different physical phenomena (resonance,
tunnelling, pair production). For tunnelling, we shall be
particularly interested in the extension of the Hartman
effect\cite{H62,Rep,PLA} from the Schr\"odinger equation to the
Dirac equation\cite{KSG01,PR03}. This would justify the renaming
of the effect to the Hartman paradox since it implies the
existence of super-luminal velocities. Within the Klein zone we
will again use the two step method described in this paper.

\newpage

\section*{APPENDIX.}
 In order to obtain the
expression of the transmitted probability for the step potential
case, we consider the following incident wave packet,
\begin{equation}
\Phi_{\I,\i}(z,t) =
\int_{p_{\m}}^{^{\infp}}\hspace*{-.2cm}\mbox{d}p\,\,\,
g(p)\,\left(\,\begin{array}{c}1\\0\\
\displaystyle{\frac{p}{E+m}}\\0
\end{array}\,\right)\, \exp[\,i\, (\,p\,z\,-E\,t\,)\,]\,\,,
\end{equation}
where $p_{\m}=\sqrt{V_{\0}(V_{\0}+2m)}$ and $g(p)$ is a real
function with a pronounced peak about the value $p=p_{\0}$ chosen
by construction such that
\begin{equation}
\int_{_{\infm}}^{^{\infp}}\hspace*{-.2cm}\mbox{d}z\,\,\,
|\Phi_{\I,\i}(z,t)|^{^{\2}}
=\int_{p_{\m}}^{^{\infp}}\hspace*{-.2cm}\mbox{d}p\,\,\,g^{\2}(p)\,\left[\,1+
\left(\frac{p}{E+m}\right)^{\2}\,\right]=1\,\,.
\end{equation}
The transmitted wave packet can then be written as
\begin{eqnarray}
\Phi_{\II}(z,t) & = &
\int_{p_{\m}}^{^{\infp}}\hspace*{-.2cm}\mbox{d}p\,\,\,
g(p)\,\,T{\pl}(0)\,\left(\,\begin{array}{c}1\\0\\\displaystyle{\frac{q}{E-V_{\0}+m}}\\0
\end{array}\,\right)\, \exp[\,i\, (\,q\,z\,-E\,t\,)\,] \nonumber \\
 & \simeq & \left[\,T{\pl}(0)\,\left(\,\begin{array}{c}1\\0\\
 \displaystyle{\frac{q}{E-V_{\0}+m}}\\0
\end{array}\,\right)\,\right]_{\mbox{\small
$p=p_{\0}$}}\hspace*{-.8cm}\exp[\,i\,q_{\0}\,z\,]\,\times
\nonumber \\ & &
\int_{p_{\m}}^{^{\infp}}\hspace*{-.2cm}\mbox{d}p\,\,\,
g(p)\,\,\exp\left[\,i\, (\,p-p_{\0})
\,\frac{p_{\0}(E_{\0}-V_{\0})}{q_{\0}E_{\0}}\,z\,-E\,t\,)\,\right]\,\,.
\end{eqnarray}
Consequently,
\begin{eqnarray}
\int_{_{\infm}}^{^{\infp}}\hspace*{-.2cm}\mbox{d}z\,\,\,
|\Phi_{\II}(z,t)|^{^{\2}} & \simeq &
\left[\,|\,T{\pl}(0)\,|^{^{\2}}\,\right]_{\0}\,
\frac{E_{\0}-V_{\0}}{E_{\0}-V_{\0}+m}\,\frac{q_{\0}E_{\0}}{p_{\0}(E_{\0}-V_{\0})}
\int_{p_{\m}}^{^{\infp}}\hspace*{-.2cm}\mbox{d}p\,\,\,g^{\2}(p) \nonumber \\
& \simeq & \left[\,|\,T{\pl}(0)\,|^{^{\2}}\,\right]_{\0}\,
\frac{2(E_{\0}-V_{\0})}{E_{\0}-V_{\0}+m}\,\frac{q_{\0}E_{\0}}{p_{\0}(E_{\0}-V_{\0})}
\, \frac{E_{\0}+m}{2E_{\0}} \nonumber \\
 & = & \left[\,|\,T{\pl}(0)\,|^{^{\2}}\,\frac{b}{a}\,\right]_{\0}\,\,.
\end{eqnarray}

\newpage

\begin{figure}[hbp]
\hspace*{-1.5cm}
\includegraphics[width=18cm, height=22cm, angle=0]{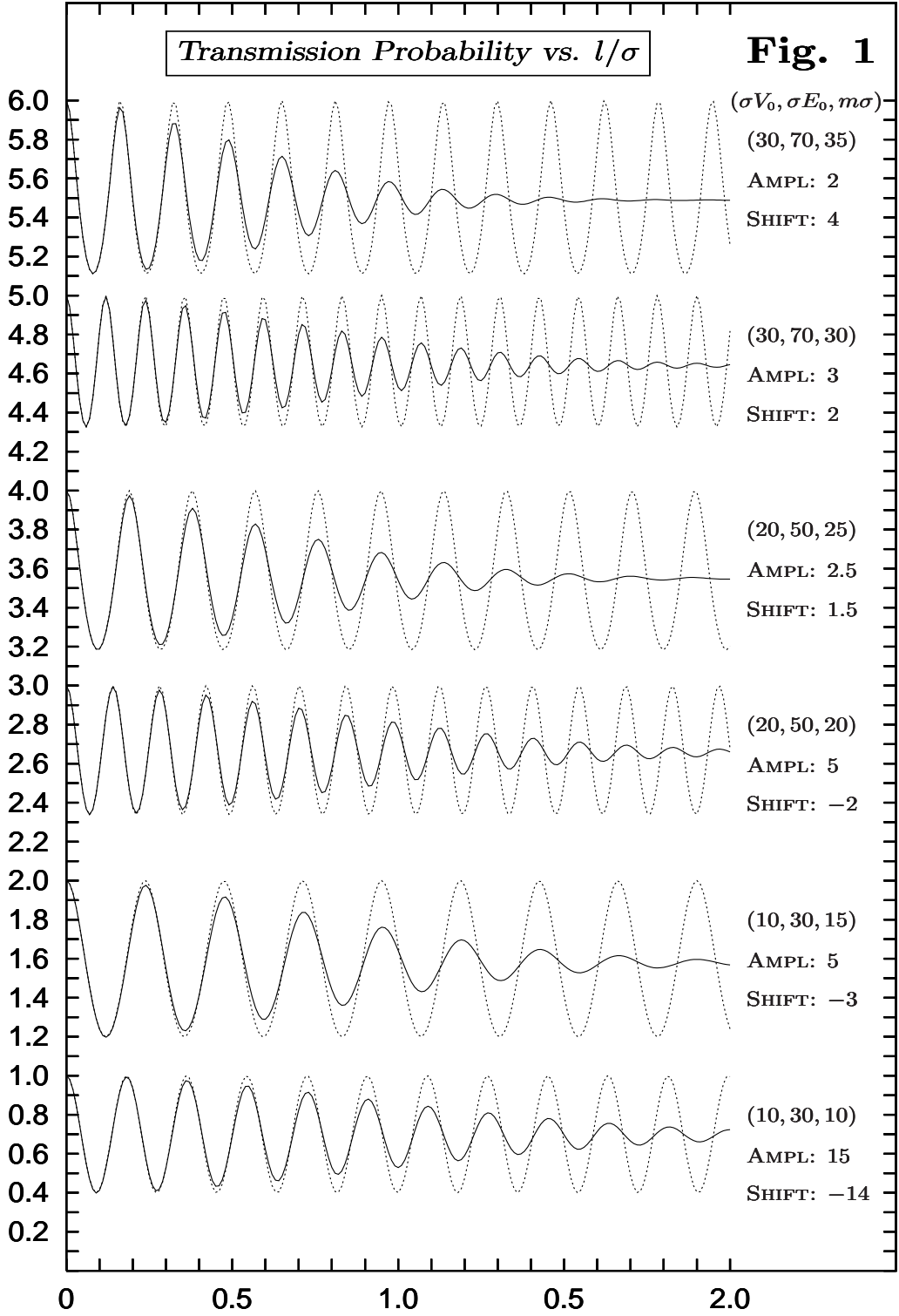}
\vspace*{-1.5cm} \caption{Barrier width dependence of the
transmission probability for different values of $\sigma V_{\0}$,
$\sigma E_{\0}$ and $m \sigma$ ($\hbar=c=1$). The transmission
probability $|T(p_{\0})|^{\2}$ (see dotted lines) shows the
typical resonance curves for plane waves. In our numerical
calculations,  performed for asymptotic times, we have used a
gaussian incoming wave packet of width $\sigma$. The transmission
probability $\sqrt{2/\pi} \,\sigma\, \int \mbox{d}p
\exp[-\sigma^{\2}(p-p_{\0})^{\2}]\,|T(p)|^{\2}$ (see solid lines)
exhibits the tendency to the constant value
$2a_{\0}b_{\0}/(a_{\0}^{\2} +b_{\0}^{\2})$ as predicted by the
multiple scattering analysis.}
\end{figure}

\end{document}